\documentclass[aps,amsmath,amssymb,showpacs]{revtex4}
\usepackage{color}
\usepackage{amssymb}
\usepackage{amsmath}%

\usepackage{amsfonts}%
\usepackage{graphicx}
\usepackage{url}

\newcommand{\e}{{\rm e}}
\newcommand {\I}{{\rm i}}
\DeclareMathOperator{\Cov}{Cov}%
\DeclareMathOperator{\Var}{Var}%
\DeclareMathOperator{\Si}{Si}

\begin{document}

\title{Self-averaging characteristics of spectral fluctuations}
\author{Petr Braun$^{1,2}$ and Fritz Haake$^1$}
\email{petr.braun@uni-due.de}
 \address{ $^{1}$Fakult\"at f\"ur Physik, Universit\"at Duisburg-Essen,
47048
 Duisburg, Germany}
\address{ $^2$Institute of Physics, Saint-Petersburg University,
198504 Saint-Petersburg,
  Russia}

\begin{abstract}
  The spectral form factor as well as the two-point correlator of the
  density of (quasi-)energy levels of individual quantum dynamics are
  not self-averaging. Only suitable smoothing turns them into
  useful characteristics of spectra. We present
  numerical data for a fully chaotic kicked top, employing two types
  of smoothing: one involves primitives of the spectral correlator,
  the second a small imaginary part of the quasi-energy.
  Self-averaging universal (like the CUE average) behavior is found
  for the smoothed correlator, apart from noise which shrinks like
  $1\over\sqrt N$ as the dimension $N$ of the quantum Hilbert space
  grows.  There are periodically repeated quasi-energy windows of
  correlation decay and revival wherein the smoothed correlation
  remains finite as $N\to\infty$ such that the noise is negligible. In
  between those windows (where the CUE averaged correlator takes on
  values of the order ${1\over N^2}$) the noise becomes dominant and
  self-averaging is lost. We conclude that the noise forbids
  distinction of CUE and GUE type behavior.  Surprisingly, the
  underlying smoothed generating function does not enjoy any
  self-averaging outside the range of its variables relevant for
  determining the two-point correlator (and certain higher-order
  ones). --- We corroborate our numerical findings for the noise by
  analytically determining the CUE variance of the smoothed
  single-matrix correlator.

\end{abstract}
 \maketitle

\section{Introduction}

Recent semiclassical work based on Gutzwiller's periodic-orbit theory
has revealed universal spectral fluctuations for quantum dynamics with
a fully chaotic classical limit
\cite{Berry87,Siebe01,Heusl07,Muell09,Haake10,Braun12b}.  However,
the present semiclassical theory leaves not satisfactorily answered
the question whether universal behavior prevails only under the
protection of suitable averages over ensembles of quantum systems
which all share the same classical limit (''$\hbar$-averages'').

We have therefore thought desirable a thorough investigation of
spectral fluctuations of individual quantum dynamics and have chosen a
kicked top without time reversal invariance for a case study. As is
well known, the spectral form factor $K(n)={1\over
  N}|\mathrm{Tr}U^n|^2$ (where $n=1,2,\ldots$ is a dimensionless
discrete time and $N$ the dimension of the Floquet matrix $U$) as
well as the two-point correlator $C(e)$ of the density of levels
(the Fourier transform of $K(n)$) need smoothing in order to
become self-averaging indicators of universal spectral
fluctuations (or absence thereof) \cite{Pande79,Prang97,Smila99}.
We have checked a certain second primitive of the form factor to
be self-averaging and faithful to the average over the circular
unitary ensemble (CUE) for times $n$ not negligibly small compared
to the Heisenberg time $N$. The correlator, a periodic function of
a quasi-energy variable $e$ conjugate to the time $n$, has
self-averaging first and second primitives inside certain
quasi-energy windows within which correlations have not yet
subsided to ''noise''. Outside those windows of correlation decay
and revival (which are tiny in width compared to the period of
$C(e)$), the noise (of order $N^{-{1/ 2}}$) overwhelms the CUE
average (order ${1\over N^2}$ for the correlator, ${1\over N}$ for
its first primitive) and self-averaging ceases to reign. We are
led to the same conclusion when smoothing the correlator by
allowing for an imaginary part Im$e>0$, the latter sufficiently
large (see below) but smaller than the mean spacing of the
eigenphases of $U$.

We represent the correlator as a descendant of a generating
function $\mathcal Z(e,\delta_+,\delta_-)$, a periodic function of
three variables. Only the behavior near $\delta_\pm=0$ determines
the correlator $C(e)$.  A primitive of $\mathcal Z$ w.r.t.~$e$
turns out smooth, self-averaging, and faithful to the CUE average
for an individual kicked top, for $\delta_\pm$ near zero and as
long as $e$ remains in the windows of correlation decay and
revival. Outside, however, there is no self-averaging. Inasmuch as
such absence of universality is irrelevant for correlator and form
factor (and certain higher-order kins, see below) one might
dispatch it as physically uninteresting. Nonetheless, the question
arises as to why previous semiclassical work has yielded the RMT
generating function without manifest necessity for ensemble
averaging. We confirm our numerical findings for the role of noise
in Sect.~\ref{sec:CUEfluct} by analytically determinating the CUE
variance of the smoothed single-$U$ correlator, building on
results of Conrey et al for higher-order generating functions
\cite{Conre07}. We thus generalize the previously known
'ergodicity' of the correlator within the CUE.

The Floquet operator $ U=\exp\left(-\I{\tau_zJ_z^2\over
    2j+1}-\I\alpha_zJ_z\right) \exp\left(-\I{\tau_yJ_y^2\over
    2j+1}-\I\alpha_yJ_y\right) \exp\left(-\I{\tau_xJ_x^2\over
    2j+1}-\I\alpha_xJ_x\right)$ of our kicked top lacks time reversal
invariance. The rotation angles were chosen as
$\alpha_y=\alpha_y=1,\alpha_x=1.1$ and the torsion strengths as
$\tau_z=10, \tau_y=0,\tau_x=4$, to bring about predominance of
classical chaos; islands of regular motion cumulatively cover less
than a single Planck cell. The angular momenta $J_{x/y/z}$ obey
$[J_x,J_y]=\I J_z$ etc.. The quantum number $j$ fixes the dimension of
the quantum Hilbert space as $N=2j+1$; our calculations involve values
of $j$ between $10^2$ and $10^4$.

\section{Generating function,
generalized correlator and form factor}
\label{sec:genf&c}

The analysis of spectral fluctuations of unitary quantum maps is
conveniently based on the generating function
\begin{align}
  \label{eq:1}
  \mathcal Z(a,b,c,d)&=\frac{1}{2\pi}\int_0^{2\pi}
         {\det(1-c\,\e^{\I\phi}U) \det(1-d\,\e^{-\I\phi}U^\dagger)
          \over
          \det(1-a\,\e^{\I\phi}U)\det(1-b\,\e^{-\I\phi}U^\dagger)}\, d\phi
                                  \,,
 \qquad |a|,\,|b|< 1,\quad c,d\in\mathbb{C}\,.
\end{align}%
  That
function \cite{Zirnb99,Braun12b} generates the two-point correlator of
the eigenphase density of $U$ as $C(e)=\textstyle{2cd\over
  N^2}\partial_c\partial_d \mathcal Z\big|_{a=b=c=d=\exp(\I{e\over
    N})}$.

The center-phase average entails interesting properties. If we set
$a=c$, two of the spectral determinants in (\ref{eq:1}) cancel. The
$\phi$-integral then simply yields $\mathcal Z(a,b,a,d)=1$, as is
easily checked by going to a complex $\phi$-plane and complementing
the integral to one over a suitable closed loop not encircling the
poles provided by the eigenphases of $U$. Similarly, $\mathcal
Z(a,b,c,b)=1$ such that we have $\mathcal Z-1= (a-c)(b-d)[\ldots ]$
with $[\ldots]$ a polynomial of order $N-1$ in both $c$ and $d$. In
fact, the quantity ${\mathcal Z-1\over (a-c)(b-d)}=[\ldots ]$ yields
the complex two-point correlator as
\begin{align}
\label{eq:3}
C(e)={2ab\over N^2}{\mathcal Z-1 \over
  (a-c)(b-d)}\Bigg|_{a=b=c=d=\e^{\I e/N}}\,.
\end{align}
The foregoing result suggests to define an algebraic kinsman  of
the generating function to be called generalized correlator,
\begin{align}
  \label{eq:4}
  \mathcal C=\Big({ab\over cd}\Big)^{N-1\over 2}\,
                    {2ab\over N^2}{\mathcal Z-1 \over (a-c)(b-d)}
\end{align}
which yields the correlator as $C(e)=\mathcal C
\big|_{a=b=c=d=\e^{\I e/N}}$, without differentiation.  Note that we
have sneaked in the factor $(ab/ cd)^{(N-1)/ 2}$ which becomes unity
for the correlator $C$; our motivation for that import will
be revealed below.

As a further consequence of the center-phase average, the
generating function and the generalized correlator $\mathcal C$ depend
on the four complex variables $a,b,c,d$ only through three independent
combinations which may be chosen as
\begin{align}
  \label{eq:5}
  ab=\e^{\I 2e/N}\,,\qquad {c\over a}=\e^{\I\delta_+/N}\qquad {d\over b}=\e^{\I\delta_-/N}\,.
\end{align}
While $\delta_\pm$ can be arbitrary complex, the variable $e$ is
restricted by $\mathrm{Im}\,e>0$. Since we loose no physically
important information by restricting ourselves to real $\delta_\pm$
that choice will be made without exception. Moreover, unless
noted otherwise, we shall take the quasi-energy $e$ as real in the
sense $\mathrm{Im}\,e\downarrow 0$.

We note in passing that the generating function $\mathcal Z$
defined above allows access to higher-order correlation functions,
through higher derivatives w.r.t.~to $c,d$ or to $\delta_\pm$
evaluated at $\delta_\pm=0$. All those functions then appear, of
course, with the single quasi-energy $e$ as the exclusive
argument. There is thus physics in the generalized correlator
$\mathcal C$ for $\delta_\pm$ near zero, for which reason we shall
check for self-averaging there as well, not just at
$\delta_\pm=0$.

We can proceed to the Fourier transform of $\mathcal
C(e,\delta_+,\delta_-)$ with respect to $e$,
\begin{align}
  \label{eq:6}
   {\textstyle\mathcal K(n,\delta_+,\delta_-)}=\int_0^{\pi N}\! \textstyle{de\over 2\pi}
                                      \,\mathcal C(e,\delta_+,\delta_-)
                                      \,\e^{-\I 2en/N}\,,
\end{align}
which we shall refer to as the generalized form factor since it
reduces to the physical form factor for $\delta_+=\delta_-=0$.

For a given spectrum $\{\e^{-\I\phi_\mu},\,\mu=1,2,\ldots N\}$ of the
Floquet operator $U$, the generating function $\mathcal Z$ can be
further evaluated by doing the center-phase average explicitly with
the help of Cauchy's theorem.  We may write the result as
\begin{align}
  \label{eq:7}
  \mathcal C(e,\delta_+,\delta_-)=  {2\over N^2} \sum_{\mu,\nu=1}^N
     {z\e^{-\I\Delta_{\mu\nu}}\over 1-z\e^{-\I\Delta_{\mu\nu}}}\,
      f_\mu(\delta_+)  f_\nu(-\delta_-)\,,\qquad \mathrm{and}\qquad
  \mathcal K(n,\delta_+,\delta_-)={1\over N}\sum_{\mu,\nu}\e^{-\I n\Delta_{\mu\nu}}
      f_\mu(\delta_+)  f_\nu(-\delta_-)
\end{align}
with
\begin{align}
  \label{eq:8}
   z=\e^{\I 2e/N}\,,\qquad
   \Delta_{\mu\nu}=\phi_\mu-\phi_\nu\,,\qquad \mathrm{and}\qquad
   f_\mu(\delta_\pm)=\prod_{\nu(\neq
  \mu)}{\sin{\Delta_{\nu\mu}+\delta_\pm/N\over 2}\over
  \sin{\Delta_{\nu\mu}\over 2}}\,.
\end{align}
We read off the period $\pi N$ for the $e$-dependence of $\mathcal C$
which allows to limit the range of $e$ to $[-\pi N/2,\,\pi N/2]$. Four
properties of the auxiliary function $f_\mu(\delta_\pm)$ and their
consequences for the generalized correlator and form factor are worth
noting: (i) For real arguments $\delta_\pm$ the function
$f_\mu(\delta_\pm)$ is real, whereupon Re\,$\mathcal C$ exclusively
parents the real correlator $R(e)=\mathrm{Re}\,C(e)$ through
Re\,${z\e^{\I \Delta_{\mu\nu}}\over 1-z\e^{\I
    \Delta_{\mu\nu}}}=-{1\over 2}+\pi\delta_{2\pi}({2e\over
  N}-\Delta_{\mu\nu})$ --- with the periodic delta functions arising
for Im\,$e\downarrow 0$ --- and analogously for the imaginary
parts which contain principal-value $1\over e$-singularities. That
property is owed to the factor $({ab\over cd})^{N-1\over 2}$ in
the definition (\ref{eq:4}) of $\mathcal C$. A fine scale of
variation in $e$ is worth being mentioned. Imagining the
eigenvalue differences $\Delta_{\mu\nu}$ ordered as they increase
from $-\pi$ to $\pi$ we can speak of a 'mean spacing'${2\pi\over
N^2}$ and the scale $\sim {1\over N}$ for $e$ .  (ii) The product
structure of $f_\mu$ reveals Fourier components
$\e^{\I\nu\delta_\pm/2N}$ for $\mathcal C$ and $\mathcal K$ with
$\nu=\pm(N-1),\pm(N-3),\ldots$. Therefore, $\mathcal C$ and
$\mathcal K$ are periodic in $\delta_\pm$ with period $2\pi N$ for
$N$ odd and $4\pi N$ for N even; moreover, the finest scale of
variation in $\delta_\pm$ is $2\pi$. (iii) Due to $f_\mu(0)=1$ we
again see the generalized correlator and form factor reduce to the
physical ones for $\delta_\pm=0$. (iv) For real
$\delta_+=-\delta_-$, $\mathcal K$ becomes non-negative while
$\mathcal C$ acquires a real (imaginary) part even (odd) in $e$.
In particular, $R(e)=R(-e)$, and therefore we need to look at the
real correlator $R(e)$ only in the interval $[0,\pi N/2]$.

In preparation of the intended use of primitives of $\mathcal C$ we
would like to point out that the explicit form of $\mathcal C$ given
in (\ref{eq:7}) is a good starting point for integrating over $e$
(recall $z=\e^{\I 2e/N}$), inasmuch as we are facing a slightly
compacted form of the partial-fraction decomposition in terms of
$(1-z\e^{-\I\Delta_{\mu\nu}})^{-1}$.  The first primitive
$\int^ede'\mathcal C(e',\delta_+,\delta_-)$ varies with the phase
$e\over N$ on the scale $1\over N^2$, the latter measuring the typical
distance between neighboring quasi-energy differences
$\Delta_{\mu\nu}$.  The real part of the first primitive is thus
piecewise constant in $e$, with steps of height $\propto {1\over N}$
and width $\propto {1\over N^2}$. For $N\gg 1$ the real part of the
first primitive will appear smooth on the scales $e\sim 1$ and
$\int^ede'\mathcal C\sim 1$. On those scales, the imaginary part of
$\int^ede'\mathcal C\sim 1$ also appears smooth, its inconspicuous
logarithmic singularities notwithstanding. Given the effective
smoothing, one expects variation of the first primitive of $\mathcal
C$ on the scale $e\sim 1$ only.

\section{CUE averages}

Checking for self-averaging and universality means comparing
spectral characteristics with their CUE averages. The generating
function \cite{Mehta04,Conre07,Braun12b}, $\big (\mathcal
Z_{\mathrm{CUE}}-1 \big)\big/[(a-c)(b-d)]= \big (1-(cd)^N
\big)\big/[(1-ab)(1-cd)]$, yields,
\begin{align}
  \label{eq:10}
  {\mathcal C}_{\mathrm {CUE}}(e,\delta_+,\delta_-)
   =\frac{\I \e^{\I e}}{N^2\sin \frac e N}\quad
     \frac{\sin\big(e+\frac{\delta_+\, +\delta_-}{2}\big)}{\sin
       \big[{1\over N}\big(e+\frac{\delta_+\,
     +\delta_-}{2}\big)\big]}
\qquad\Longrightarrow\qquad
  {\mathcal C}_{\mathrm {CUE}}(e,\delta,-\delta)=C _{\mathrm {CUE}}(e)={\e^{{\I} 2e}-1 \over 2N^2\sin^2(e/N)}\,.
\end{align}
The generalized correlator depends on the phases $\delta_\pm$ only
through their sum $\delta_+\,+\delta_-$; that property is shared
by the generalized form factor (\ref{eq:6}). The limit $N\to
\infty$ with ${e\over N}\to 0$ gives $C_{\rm{CUE}}(e)\to
C_{\rm{GUE}}(e)={\e^{{\I}
    2e}-1 \over 2e^2}$ and correspondingly $\mathcal
Z_{\rm{GUE}}(e,\delta_+,\delta_-)$ and $\mathcal
C_{\rm{GUE}}(e,\delta_+,\delta_-)$. The two
  determinants in the denominator of (\ref{eq:1}) have their zeros
  along the $\phi$-axis shifted by $2e$ whereas those in the numerator
  have relative shift $2e+\delta_+\,+\delta_-$. It is these two shifts
  which are the arguments in the factors of the CUE generalized
  correlator.

\section{Smoothing by integration}
\label{sec:primitives}

The first primitive of the real correlator $R=\mathrm{Re}\,C$ can be
defined as $R^{(1)}(e)=\int_e^{N\pi /2} de'R(e')$ with the reference
point at the right border of the principal interval of $e$.  It is
close to zero for all $e$ comparable to $N$; when $e$ tends to $+0$
the first primitive of the physical correlator with
$\delta_+=\delta_-=0$ tends to the universal limit $-\pi/2$ for any
spectrum. The first primitive experiences a discontinuity at $e=0$ due
to the diagonal terms in the sum (\ref{eq:7}). When $\delta_\pm=0$
and, generally, when $\delta_+=-\delta_-$ we have $R^{(1)}(0)=0$ such
that the first primitives with the reference point at zero and at
$N\pi/2$ coincide, up to the sign. These properties can be understood
considering that every physical pair of eigenphases makes one and one
only contribution to $R^{(1)}(0)$.

\begin{figure}[tbh]
\includegraphics[width=0.47\textwidth]{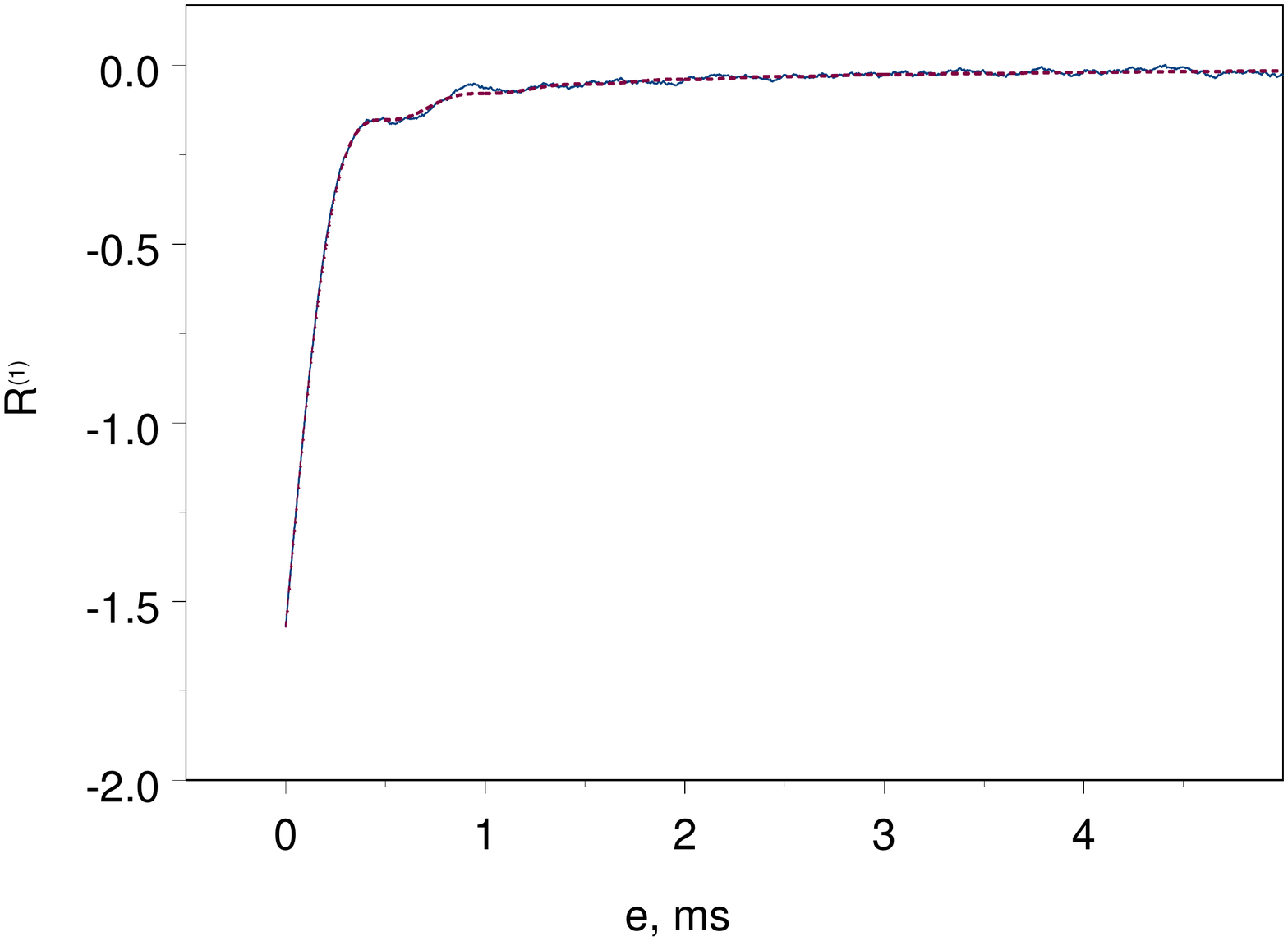}
\hfill
\includegraphics[width=0.47\textwidth]{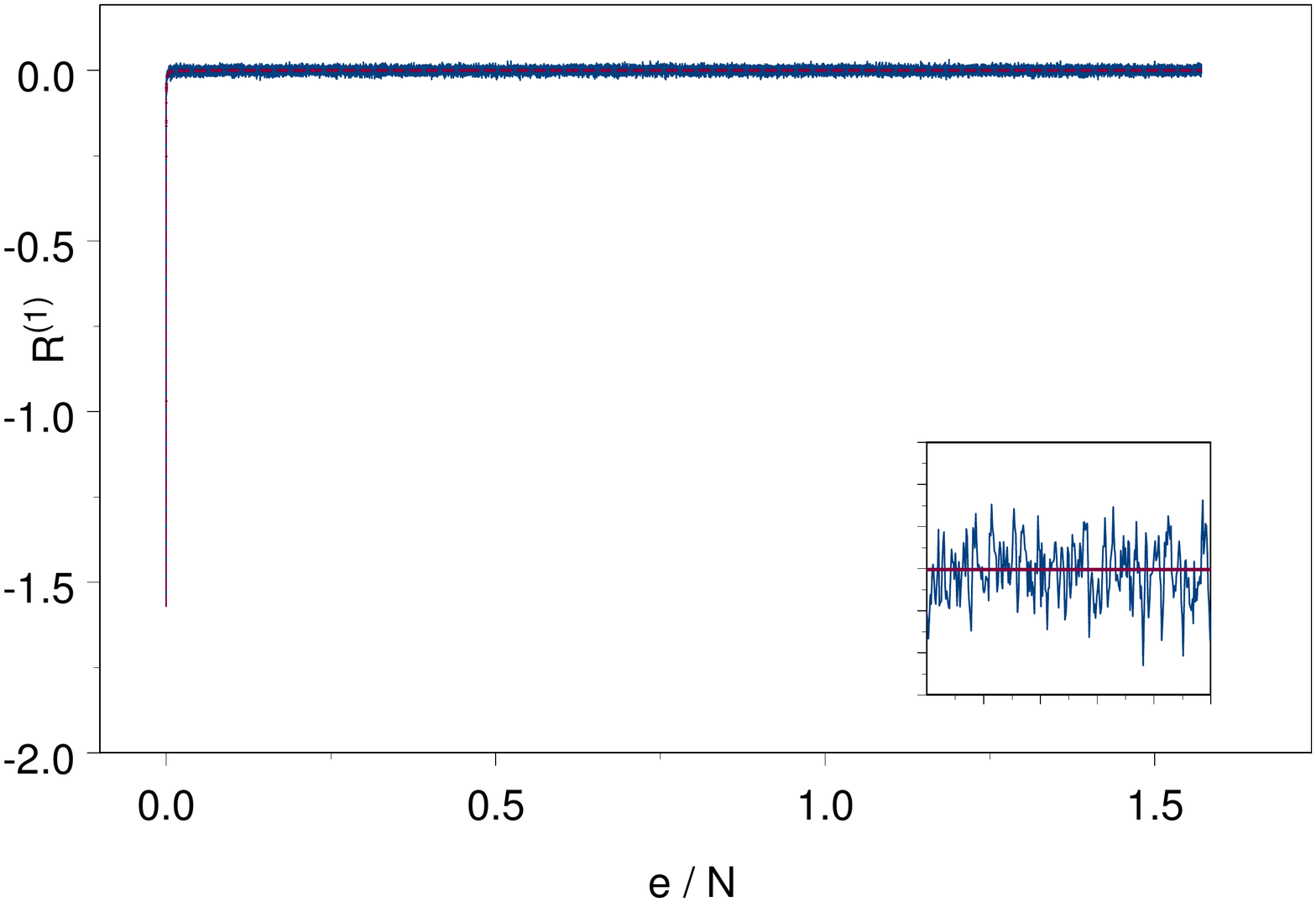}
\parbox[t]{0.47\textwidth}{\caption{Kicked top, $j=9600$. First
    primitive  $R^{(1)}(e)$ with $\delta=0$ practically coincides
    with RMT (dashed)} 
\label{1stprim_a}}
\hfill
\parbox[t]{0.47\textwidth}{\caption{Same but for large $e$. Only noise
    remains once $e$ exceeds a few ms. Blowup of noise stretch in
    inset}\label{1stprim_b}}
\end{figure}

The first primitive is known to be self-averaging and faithful to
random-matrix theory for fully chaotic dynamics \cite{Haake10}, as
illustrated for the kicked top in Fig.~\ref{1stprim_a}; the phase
${e\over N}$ is reckoned there in units of the mean level spacing
(ms) ${2\pi\over N}$, such that the mean level spacing in terms of
the quasi-energy $e$ is $2\pi$.  However, self-averaging and
universality reign only for $e$ in a range of a few ms, in the
region of noticeable level-level correlations. --- Outside that
region, in particular for $e\sim N$, $R^{(1)}(e)$ fluctuates
around zero without systematic $e$-dependence, see
Fig.~\ref{1stprim_b}; the black ribbon of constant width consists
of irregular fluctuations; the inset shows a blow-up. The noise
amplitude turns out to decay as $1/\sqrt{N}$ with $N\to \infty$
(accuracy better than 2\% for absolute deviation from mean, over
two decades of $N$). Both the CUE average,
$R^{(1)}_{\mathrm{CUE}}(e)\propto N^{-1}$ and the difference
between the CUE and GUE correlators are overwhelmed by the noise.
It follows that a single spectrum does not allow to distinguish
the infinite-$N$ GUE correlator from the finite-$N$ CUE one, the
periodicity of the latter apart. --- The overall behavior of
$R^{(1)}(e)$ for individual kicked tops is consistent with the
'ergodicity' of the correlator within the CUE
\cite{Pande79,Haake10}, to be looked at more closely in
Sect.~\ref{sec:CUEfluct}.

As mentioned above, the first primitive appears smooth on the
physically most interesting scales. On the other hand, if one wants to
strictly ban all singularities one may focus on the second primitive
of $\mathcal C$ which is continuous in the phase $e$.  In that vein we
define
\begin{align}
  \label{eq:9}
  \mathcal C^{(2)}(e,\delta_+,\delta_-)=\int_0^ede'\int_{N\pi\over 2}^{e'}de''
   \mathcal C(e'',\delta_+,\delta_-)
   \qquad \mathrm{and}\qquad
   \mathcal K^{(2)}(n,\delta_+,\delta_-)=\sum_{n'=1}^n \sum_{n''=1}^{n'}
    \mathcal K(n'',\delta_+,\delta_-)\,.
\end{align}

\begin{figure}
[here]
\includegraphics[scale=0.223]{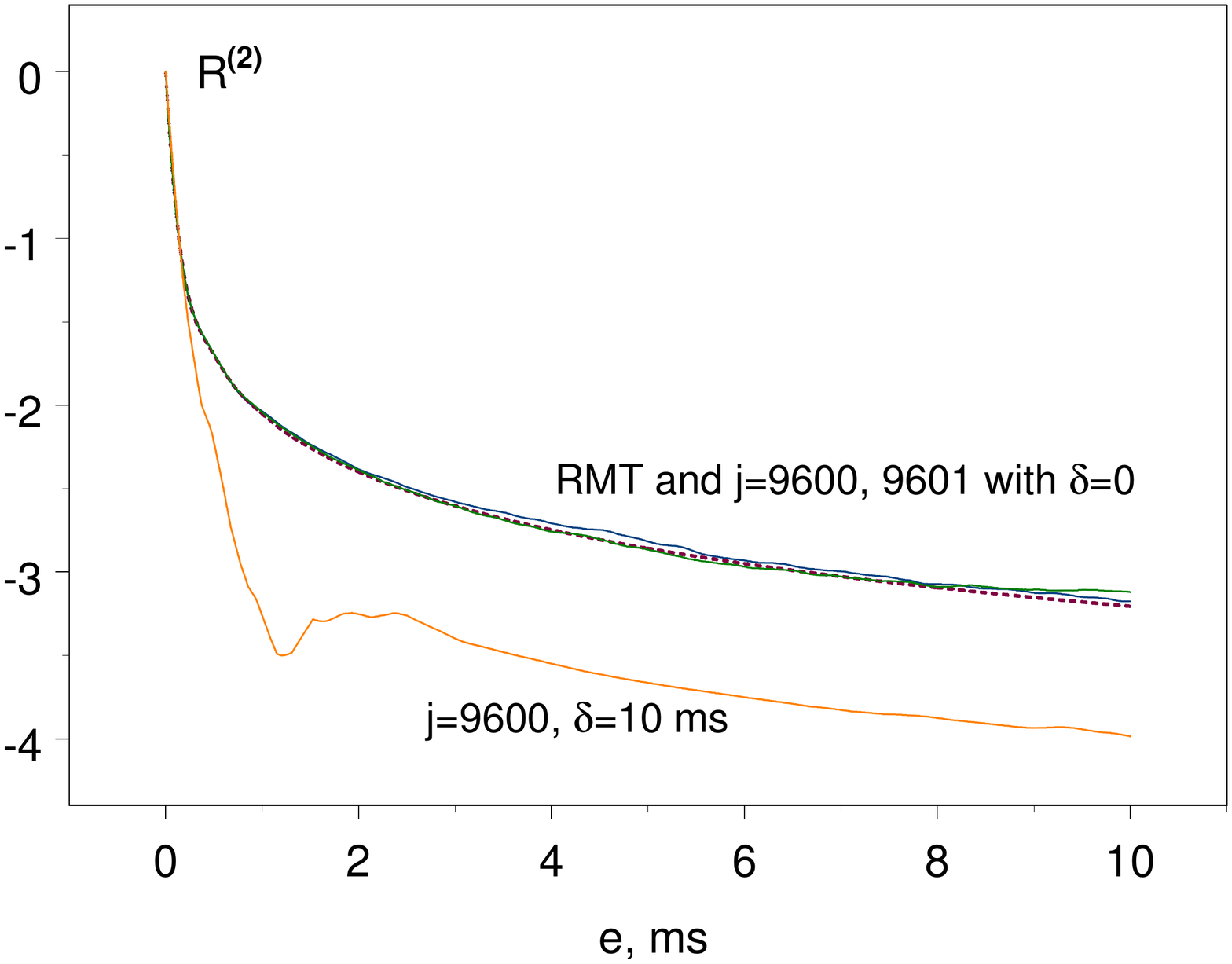}
\hspace{-1.2cm}
\includegraphics[scale=0.223]{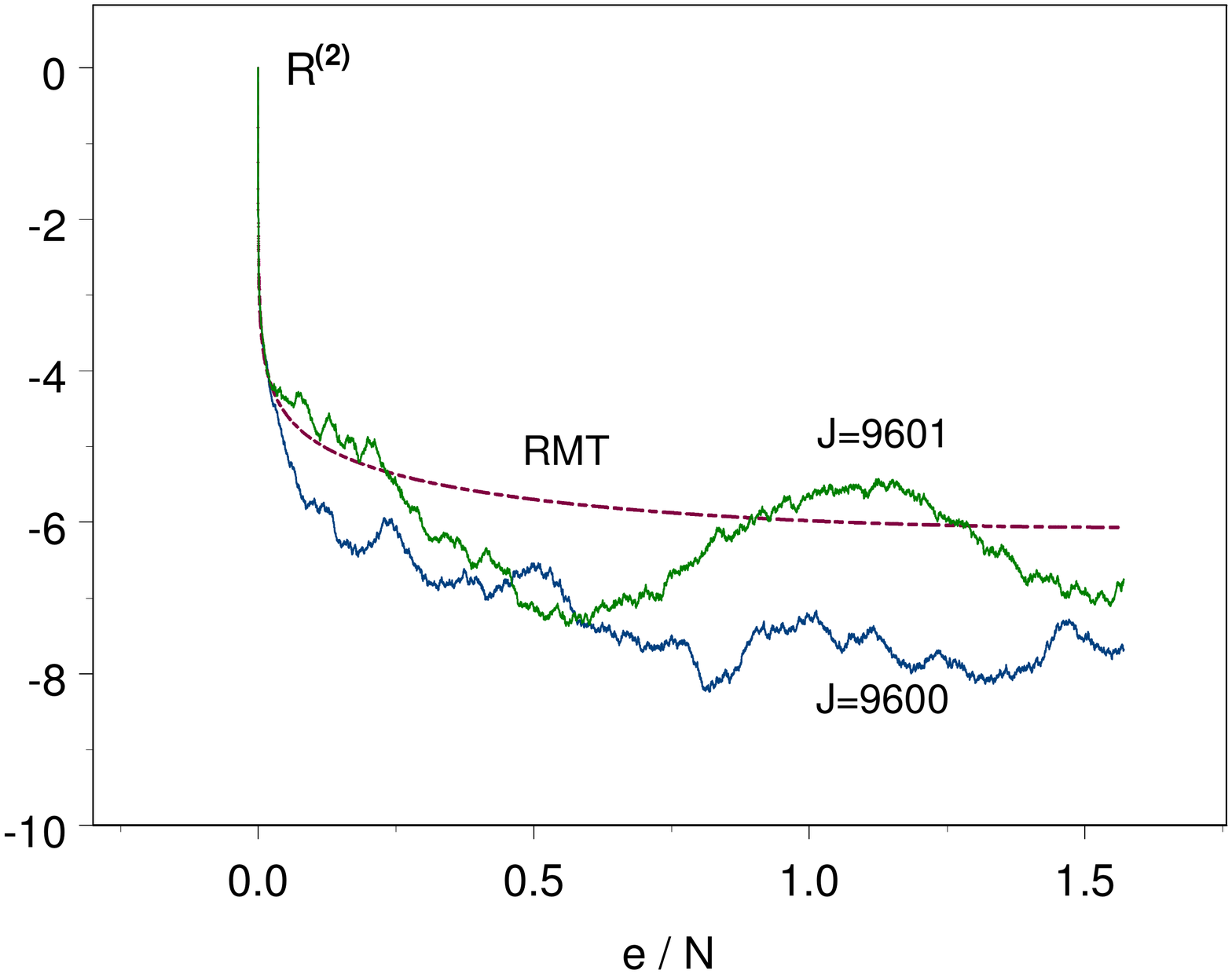}
\hspace{-1.2cm}
\includegraphics[scale=0.223]{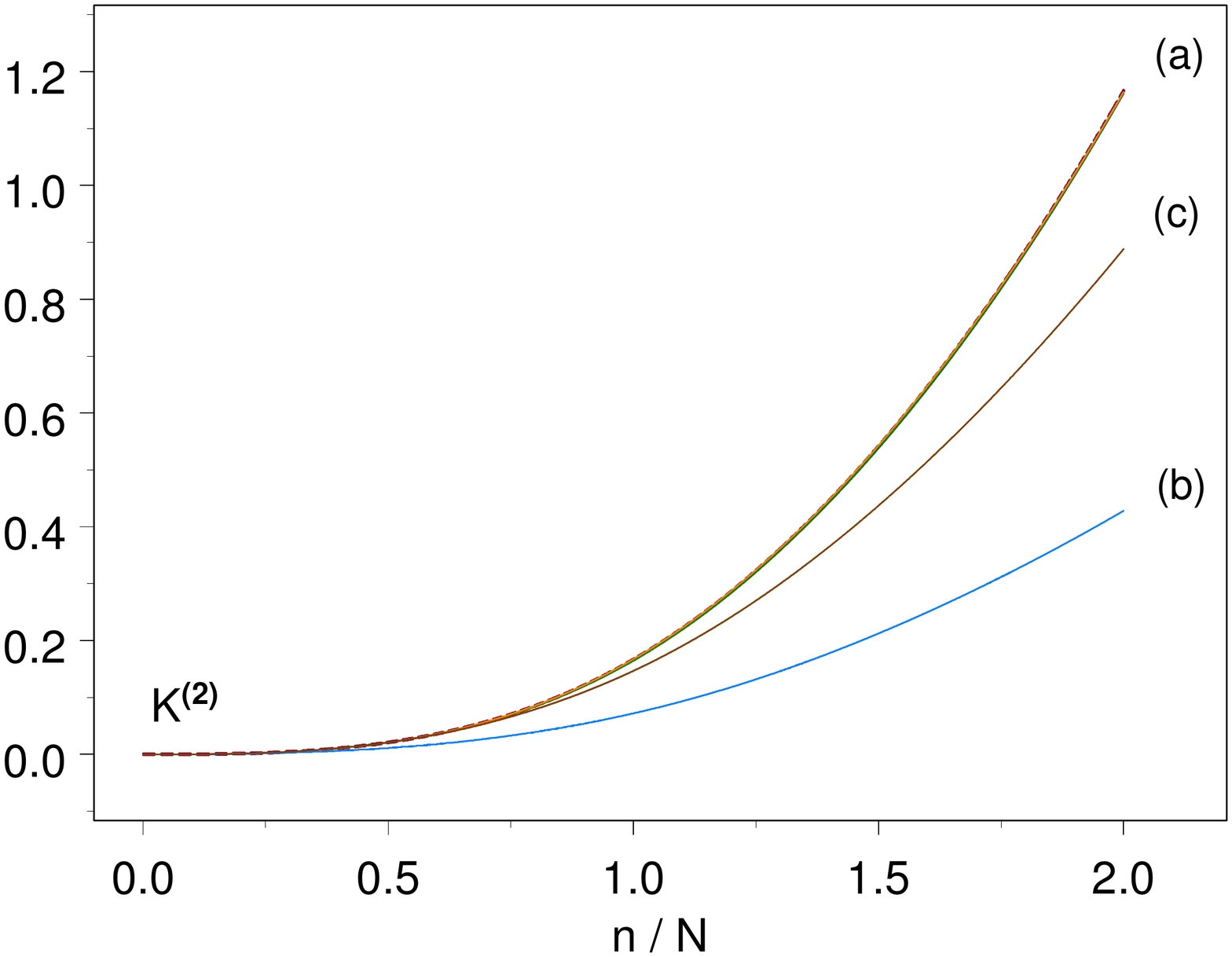}
\parbox[t]{0.32\textwidth}{\caption{self-averaging Re$\mathcal C^{(2)}$ of kicked\\
  top for $e\ll N,\,\delta=0$. Violation of\\ universality for
  $\delta=10$ms}\label{fig:gcLS}}
\parbox[t]{0.32\textwidth}{\caption{$R^{(2)}$ noisy and
    non-universal\\ for large $e$}\label{fig:gc}}
\parbox[t]{0.32\textwidth}{\caption{(a) self-averaging $\mathcal K^{(2)}$
    of \\kicked top for $\delta=0$ and $\delta=1$ms.\\(b,c) No
    universality for $\delta=10$ms}\label{fig:gff}}
\end{figure}

Our numerical studies of the second primitives reveal
self-averaging and fidelity to RMT only for $|\delta_\pm|$ at most
of order unity and for $e$ within the windows of correlation decay
(around $|e|=0 \mod N\pi$) already met with in $R^{(1)}(e)$.
The
self-averaging of $R^{(2)},\, K^{(2)}$, together with the
mentioned limitations, is revealed in Figs.~\ref{fig:gcLS},
\ref{fig:gc}, and \ref{fig:gff}.

In particular, Fig.~\ref{fig:gcLS} depicts a narrow bundle of
three curves for the second primitive of the real correlator
$R^{(2)}(e)$, one representing the CUE average and the other two
pertaining to the kicked top with $j=9600$ and 9601 in the range
of $e$ between zero and 10 ms. Throughout that range, the three
curves differ from one another by but hardly noticeable amounts.
It is to be noted that for $e$ equaling 5 ms the CUE correlator
$R_{\mathrm {CUE}}(e)$ has decayed to a practically negligible
level $\sim {1\over (10\pi)^2}$. A fourth curve in
Fig.~\ref{fig:gcLS} pertains to the second primitive of the real
part of the generalized correlator $\mathcal C(e,\delta_\pm)$ with
$\delta_+=-\delta_-=\delta$ equaling 10 mean spacings. That curve
strongly deviates from the CUE prediction (which as already
mentioned does not depend on $\delta$) and signals
non-self-averaging behavior.

For larger values of $e$ the relative differences grow and signal
non-universality, see Fig.~\ref{fig:gc}. However, that growth is of no
importance for two reasons. First, outside the phase window of
correlation decay (and revival; note the periodicity with period $\pi
N$) only weak noise remains for $R^{(1)}(e)$. Second, the second
primitive grossly exaggerates all large-$e$ structures, turning the
$1\over e^2$ decay of $R_{\mathrm {CUE}}(e)$ into a logarithmic one.

The same salient message is signaled in Fig.~\ref{fig:gff} for the
second primitive of the form factor as a function of the scaled
discrete time $n\over N$. Three curves (a), unresolved from one
another, refer to the CUE and the kicked top for $j=9600$ with
$\delta_+=-\delta_-=\delta$ equal to zero and a single mean
spacing indicate excellent self-averaging and universality. Two
further curves pertain to the kicked top with $\delta=10$ ms, one
(b) for $j=9600$ and the other (c) for $j=9601$. Universality
would require both curves (b,c) to coincide with the triple (a)
and is grossly violated.  The noisy small-time behavior of
$\mathcal K^{(2)}(n,\delta,-\delta)$, invisible on the scale of
Fig.~\ref{fig:gff}, corresponds to the large-$e$ noise in
$\mathcal C^{(2)}$.

\section{Smoothing by complex quasi-energy}
\label{sec:Ime}

\begin{figure}
[here]
\begin{center}
\includegraphics[
natheight=9.066000in, natwidth=11.733800in,
 height=2.4502in,
width=3.2166in
]%
{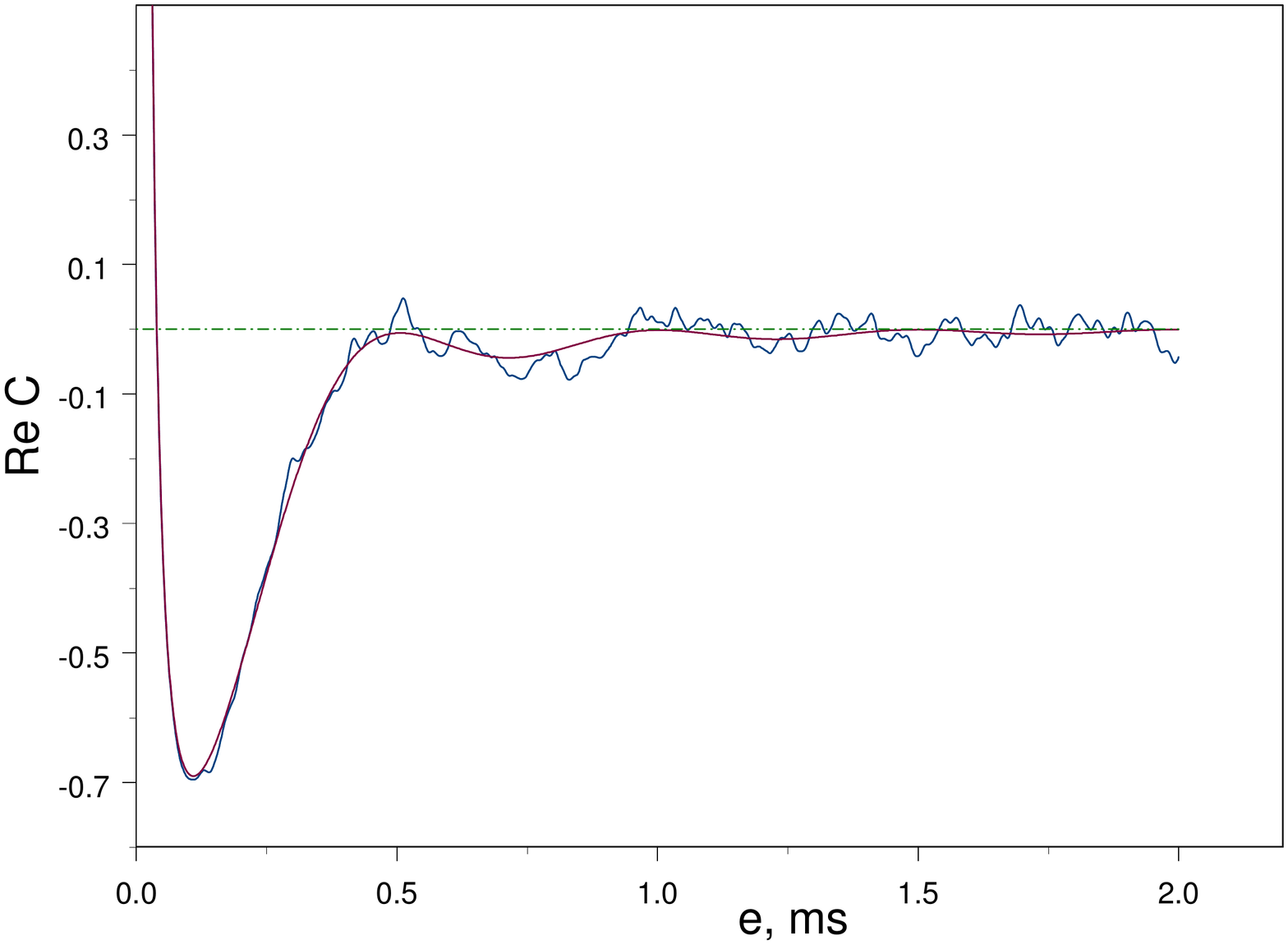}%
\caption{Physical correlator $R=\mathrm{Re}\,C$ of a kicked top
  spectrum, $j=9600$.  Regularization by imaginary part
  $\mathrm{Im}\,e=0.01$ ms. Smooth curve shows CUE prediction.}
\label{9600}
\end{center}
\end{figure}

In Fig.~\ref{9600} we show the real correlator
$R(e)=\mathrm{Re}\,C(e+\I\eta)$ where $\eta$ is a small
imaginary addition to the real quasi-energy $e$. To smooth away the
delta peaks arising for vanishing $\eta$ that imaginary part must
be larger than the finest scale of variation in $e$, seen to be $\sim
{1\over N}$ at the end of Sect.~\ref{sec:genf&c}. On the other hand,
we want to keep $2\eta\over N$ smaller than the mean spacing
$2\pi\over N$ of the eigenvalues of $U$, in order not to tamper with
the $e$ dependence on that latter scale. These restrictions are
respected in Fig.~\ref{9600}. Just as for smoothing by integration we
see excellent agreement with the CUE average within the principal
window of correlation decay but noise dominance outside: for $e\over
N$ not small the CUE average scales as $1\over N^2$ while the standard
deviation is found to decay only like $1\over \sqrt N$ as $N$ grows at
fixed $\eta$.

\section{CUE fluctuations}
\label{sec:CUEfluct}

In order to  better understand  the noise in single-top
correlators and the systematic non-self-averaging in the
generalized correlator, we here look at fluctuations of the
single-$U$ correlator throughout the CUE. We start with some
numerical findings and shall afterwards proceed to discussing an
explicit analytic result for the CUE variance.

\subsection{Numerical results}
\label{sub:num} 
\begin{figure}[here]
\includegraphics[width=0.63\textwidth]{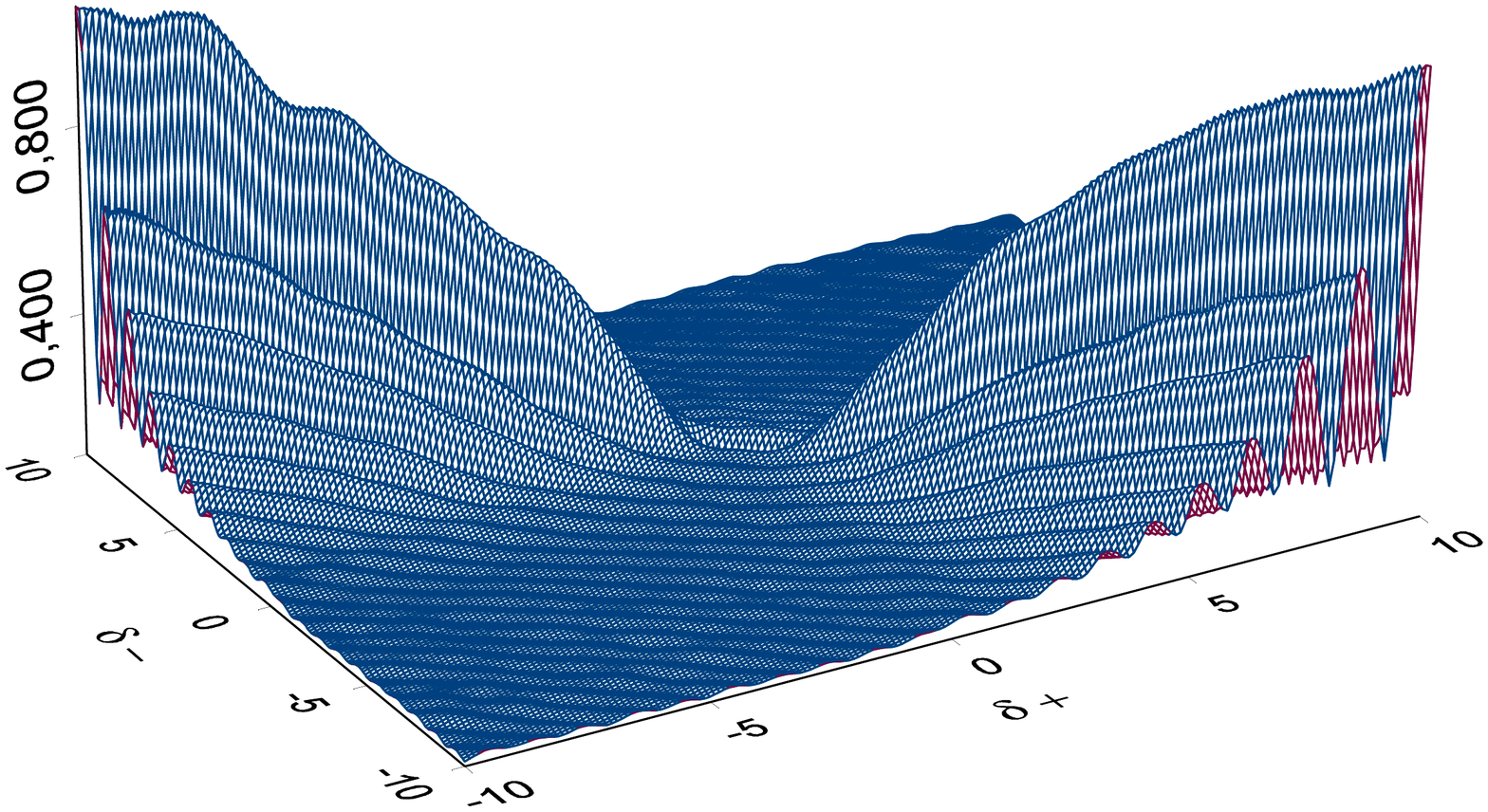}
{\caption{Absolute deviation
    $\left\langle\left | \mathrm{Re}\mathcal
        C^{(2)}-\mathrm{Re}\mathcal C^{(2)}_{CUE}
      \right|\right\rangle$ of the second primitive of the generalized
    correlator from its CUE expectation value at fixed $e=1$ ms. The
    result is an average over 13937 spectra of $201\times 201 $ random
    CUE matrices. }\label{fig:3D}}
\end{figure}

Fig.~\ref{fig:3D} further conveys the absence of
self-averaging and universality in the generalized correlator (and
thus the generating function) outside the physically relevant
range of its variables. The 3D plot in Fig.~\ref{fig:3D} shows the
mean absolute deviation $\langle|\mathrm{Re}\mathcal C^{(2)}-
\mathrm{Re}\mathcal C^{(2)}_{\mathrm{CUE}}|\rangle$, the mean
$\langle\cdot\rangle$ taken over 13937 spectra of $201\times 201$
matrices randomly drawn from the CUE (with the algorithm described
in \cite{Zyczk94}), as a function of $\delta_+$ and $\delta_-$
with $e=1$ ms; the range captured for $\delta_\pm$ is chosen as 20
ms, symmetric about zero. The line $\delta_++\delta_-=0$ is
obviously distinguished: Thereon, the mean absolute deviation is
small only in a narrow interval near $\delta_+=\delta_-=0$ while
outside a high ridge arises and signals large fluctuations of
$R^{(2)}$ within the CUE; a single unitary matrix, either for a
kicked top or drawn at random from the CUE, would entail even
stronger and noisier absolute deviation and thus reveal
non-self-averaging. Perpendicular to the line
$\delta_++\delta_-=0$ there is less drama: The mean absolute
deviation undergoes oscillations, roughly with the finest scale of
variation allowed by the Fourier series for $f_\mu(\delta_\pm)$,
and decay with increasing $|\delta_+-\delta_-|$. Even though less
flagrant and not obvious from Fig.~\ref{fig:3D},
non-self-averaging is also incurred along the line
$\delta_+-\delta_-=0$ with increasing distance from the center
$\delta_+=\delta_-=0$; already a few mean spacings beyond the
center, relative mean absolute deviations of order unity become
typical.

When accumulating the many spectra making up Fig.~\ref{fig:3D} we
have observed slow convergence, the slowness due to 'rare extreme
events'; a single odd spectrum can significantly alter the partial
average previously accumulated.  We suspect that such extreme
events are spectra with exceptionally close nearest-neighbor pairs
of levels; for such spectra the coefficients $f_\mu$ in the
spectral representation (\ref{eq:7}) become almost singular and
thus give rise to huge deviations from the universal limit and
flagrant absence of self-averaging as soon as
$|\delta_+-\delta_-|$ exceeds a few mean spacings.  That suspicion
is nourished by an analogous investigation of the circular
orthogonal ensemble where weaker level repulsion makes this
scenario more probable; as demonstrated in Section
\ref{numercuecoe} of the Appendix we found even slower convergence
and orders-of-magnitude larger variance of the generalized
correlator with the same $\delta_\pm\ne 0$ than for the CUE.
Similar calculations for the circular symplectic ensemble with the
level repulsion even stronger than in CUE, demonstrated opposite
tendencies.

\subsection{Analytic results}
\label{anal}

All of the foregoing numerical results are supported and in part
generalized by the CUE variance of the correlator smoothed by
either an imaginary part of the quasi-energy $e$ or by
integration. That variance can be extracted from the CUE-averaged
combination of eight spectral determinants $\big\langle{DDDD\over
DDDD}\big\rangle_{\mathrm{CUE}}$ where  each $D$ has its own
quasi-energy argument\cite{Conre07}; the result has yet to be
averaged over the relative central phase of the two $DD\over DD$
ratios. We shall here discuss the final expression for the
variance which generalizes Pandey's ergodicity of the correlator
within the CUE \cite{Pande79,Haake10}. For details see Appendix,
Section \ref{RMTcovvar}.

{\bf Complex quasi-energy:} We first turn to the generalized
correlator smoothed by an imaginary part $\eta$ added to the real
quasi-energy $e$. The pair $\delta_\pm$ will be left real and even
restricted as $\delta_{+}=-\delta_{-}=\delta $. The latter special
case deserves special attention due to the strong fluctuations at
large $\delta$ seen in Fig.~\ref{fig:3D} while the CUE average is
$\delta$-independent. The variance
$\mathrm{Var_{CUE}}\big(\mathrm{Re}\,\mathcal
C(e+\I\eta,\delta)\big)$ is given by a somewhat cumbersome
expression; we show only the leading term of its expansion in
powers of $\eta$,
\begin{equation}
  \mathrm{Var_{CUE}}\big(\mathrm{Re}\,\mathcal C\left(
    e+\I\eta,\delta\right)\big)  \sim{1\over 2 N\eta}
   \left(
     1-\frac{\sin^{2}\frac{\delta}{2N}}{\sin^{2}\frac{e}{N}}\right)^2
  \left(  1-\frac{\sin^{2}e}{N^{2}\sin^{2}\frac{e}{N}}\right)
  \big(1+\eta f(e,N)+\ldots\big)
\label{vargencorrsmallim}
\end{equation}
with  $f(e,N)$ finite for all $e$ and $N\to \infty$ \footnote{This
equation can be regarded as the result of replacing
$\delta(e_1-e_2)$ in the covariance of correlators  with real
energy arguments (\ref{recovar_delta0}),(\ref{recovar_delta_ne_0})
by the ``smoothed'' delta $\eta/\pi[(e_1-e_2)^2+\eta^2]$,
neglecting all other terms and then setting $e_1=e_2=e$.}.
 In particular, the variance of the physical correlator,
$\mathrm{Var_{CUE}\,Re}\big(C(e+\I\eta)\big)$, arises for
$\delta=0$ and is seen to be $\propto {1\over N\eta}$. For it to
become small for large $N$ the smoothing imaginary part $\eta$
must, as already argued in Sect.~\ref{sec:Ime}, be larger than the
minimum scale of variation in $e$, that is ${1\over N}$, and small
compared to the mean spacing, that is $2\pi$ in the units used for
$e$. These restrictions are respected for $\eta \propto
N^{-\alpha}$ with $0<\alpha<1$ and then the variance of the
physical correlator is $\propto N^{-1+\alpha}$, small indeed and
suggesting self-averaging of the correlator. On the other hand, we
find confirmed the numerical finding that outside the windows of
correlation decay and revival where $e\over N$ is of order unity,
the fluctuations overwhelm the mean; in fact the standard
deviation exceeds the mean already at much smaller energies,
namely when $e\gtrsim \sqrt[4]{\eta N}$.

Even more drastic is the non-self-averaging of the generalized
correlator when $e$ ranges within the windows of correlation decay
and revival while $\delta\sim  N^\beta,\quad 0<\beta\le 1$; the
variance then becomes $\sim N^{4\beta-1+\alpha}$.

A bit surprising is the smallness of the generalized correlator
variance ($\sim{1\over N\eta}=N^{-1+\alpha}$) when both $e$ and
$\delta$ are of order $N$, but there again fluctuations overwhelm
the mean. At any rate, the region $\delta\gg1$ is devoid of
physical interest.

\begin{figure}
[here]
\begin{center}
\includegraphics[
natheight=4.914000in,
natwidth=8.007000in,
height=2.0000in,
width=3.2000in]%
{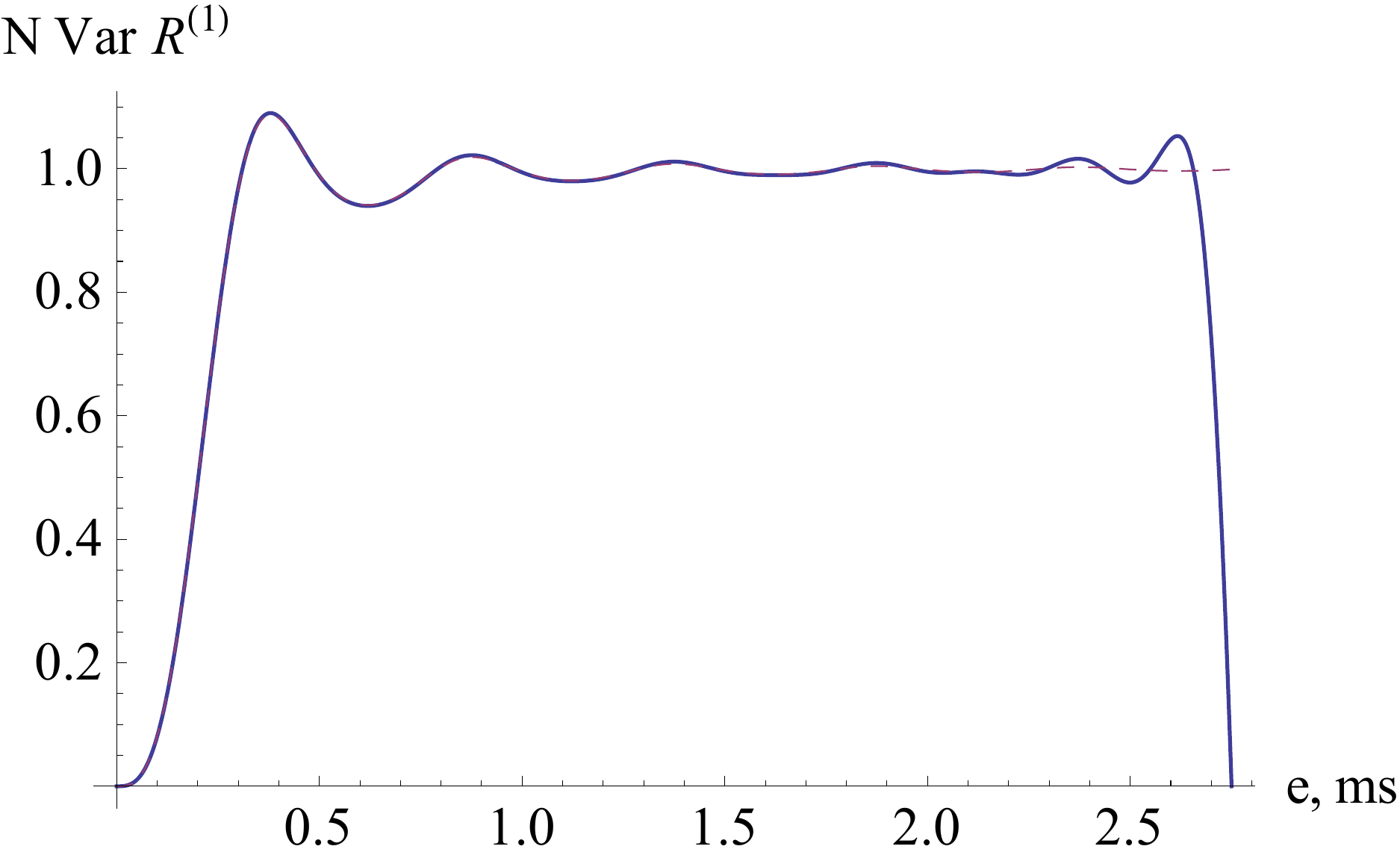}
 \caption{Variance of the first primitive of the
physical correlator
  with $N=11$, exact (full line) and large--$N$ asymptotics (dashed)}%
\label{varR1_E}%
\end{center}
\end{figure}

{\bf Primitives:}  Proceeding to smoothing by integration we note the
variance of the first primitive $R^{(1)}(e)$ at Im\,$e\downarrow 0$,
\begin{align}
\mathrm{Var_{CUE}} R^{\left(  1\right)  }(e)  & =\int_{0}^{e}dy\,\left[
\frac{\sin4y}{N^{3}\sin^{2}\frac{2y}{N}}\ln\frac{\sin^{2}\frac{e-2y}{N}}%
{\sin^{2}\frac{e}{N}}\,+\frac{\pi}{N}\Bigg(  1-\frac{\sin^{2}y}{N^{2}\sin
^{2}\frac{y}{N}}\Bigg)  \right. \label{var_delta0}\\
& \left.  +\frac{\left(  e-y\right)  \left(  4N\cos4y-2\cot\frac{2y}{N}%
\sin4y\right)  }{N^{4}\sin^{2}\frac{2y}{N}}-\frac{1}{Ny}\ln\frac{\left(
e-2y\right)  ^{2}}{e^{2}}\right]  .\nonumber
\end{align}
The variance is exactly zero at $e=0$ and $e=N\pi/2$ which is
understandable since $R^{\left( 1\right) }(e)$ of any spectrum has
the same value at these points. Away from the end points the
variance steeply rises to approximately $1/N$. It is not difficult
to extract the large-$N$ asymptotics within the principal window
of correlation decay and revival where ${e\over N}\ll 1$,
\begin{align}
\mathrm{Var_{CUE}} R^{\left(  1\right)  }(e)  & \sim\frac{1}{N}\left[
1-\frac{1}{2}\cos4e-\frac{\sin4e}{8e}-2e\,\Si 4e\right.
\label{var_delta0_appro}\\
& \left.  +\pi\left(  e+\frac{\sin^{2}e}{e}-\Si 2e\right)
+\int_{0}^{e}\frac{\left(  \sin4y-4y\right)  \log\left\vert 1-\frac{2y}%
{e}\right\vert }{2y}dy\right] \nonumber\,;
\end{align}
here $\Si x=\int_{0}^{x}\frac{\sin t}{t}dt$.
   In Fig.~\ref{varR1_E},  both
the exact and the asymptotic form of $\mathrm{Var_{CUE}} R^{\left(
    1\right) }(e)$  are depicted for $N=11$.
Both curves practically agree, except for $e$ in the immediate
neighborhood of the half period, where the large-$N$ asymptotics
cannot fall back to 0. The close agreement of the two curves in the
figure is in fact amazing since the large-$N$ asymptotics is derived
only for small $e\over N$. The scaling with $N$, $\mathrm{Var_{CUE}}
R^{\left( 1\right) }(e) \sim {1\over N}$, is manifest in the figure.

The overwhelming of the mean by fluctuations for large $e$,
${e\over
  N}\sim 1$, is found again, here for the first primitive $R^{(1)}(e)$
such that the ratio of standard deviation and mean is of the order
$\sqrt N$.

To characterize the CUE fluctuations of the generalized correlator
we finally comment on the variance of Re\,$\mathcal
C^{(1)}(e,\delta)$ where again $\delta=\delta_+=-\delta_-$. We
sketch the explicit result only for the $e$-window of correlation
decay and revival, ${e\over N}\ll 1$,
\begin{eqnarray}%
\label{varexpanded}%
 \mathrm{Var} \mathrm{Re}\, C^{(1)}(e,\delta)\approx
\big(\mathrm{Var}
R^{(1)}(e)\big)\left[1+N^2\sin^2\frac{\delta}{2N}F_1(e)\right.%
+\left . N^4\sin^4\frac{\delta}{2N}F_2(e)\right]
\end{eqnarray}
where $F_1,F_2$ are some $N$-independent functions of $e$. The
most important conclusion is negligibility of fluctuations in the
same sense as for $\delta=0$, as long as $\delta$ remains of order
unity. The extreme case of $\delta\sim N$ has $ \mathrm{Var}
\mathrm{Re}\, C^{(1)}(e,\delta)\sim N^3$ and thus no
self-averaging. Even the less excessive growth $\delta\sim
N^\beta$ with $\beta<1$ gives $ \mathrm{Var}
\mathrm{Re}\,C^{(1)}(e,\delta)\sim N^{4\beta-1}$ and thus loss of
self-averaging for $\delta>{1\over 4}$. Indeed, in our numerical
calculations the largest $N$ was of the order $10^{4}$ with
$N^{1/4}$ $\sim10$; the break-up of self-averaging could thus be
expected (and was observed) at $\delta\gtrsim2\pi=1$ ms.

\section{Discussion and outlook}

We have found fidelity to RMT of the first and second primitives
of spectral form factor $K(n)$ and real two-point correlator
$R(e)$ for an individual kicked top in periodically (period
$\propto N$) repeated windows of correlation decay and revival for
the quasi-energy variable $e$.  In between these windows
correlations are so weak as to be negligible for large $N$. System
specific noise in the $e$-dependence is found to have the order
${1\over \sqrt N}$ and thus to be negligible as well.
Interestingly, however, the noise overwhelms the CUE average of
the correlator in between the windows of correlation decay and
revival. In fact the noise is strong enough to preclude
distinguishability of the CUE and GUE prediction for a single
spectrum. The same behavior is found when the correlator is
smoothed by an imaginary addition $\I\eta$ to the real
quasi-energy $e$, provided $\eta$ is (i) large enough to 'iron
out' the singularities in the $\eta\downarrow 0$ correlator and
(ii) small enough to not noticeably attenuate $R(e)$.  On the
other hand, we find the underlying generating function $\mathcal
Z(e,\delta_+,\delta_-)$ self-averaging and universal only under
the additional restriction of $|\delta_\pm|$ no larger than a few
mean spacings.

Our results for the CUE variance of the correlator $R(e)$, the
latter smoothed by suitable $\eta$ or integration, mean that in
the large-$N$ limit any CUE matrix can with overwhelming
probability (for $N\to\infty$ with probability one) be expected to
have a smoothed correlator equaling the CUE average. Likewise, by
working out the smoothed correlator for a single top (or any other
fully chaotic map) with large $N$, one has overwhelming
probability to get the CUE average. Exceptions are possible but
will in practice not be met with.

We would like to comment on the status of previous semiclassical
work involving quadruples of classical pseudo-orbits from narrowly
packed bunches. Such work has resulted in universality of
$\mathcal Z(e,\delta_+,\delta_-)$, without restriction for the
independent variables and without manifest necessity of any
ensemble average\cite{Muell09,Braun12b}. That periodic-orbit
approach involves a 'perturbation series' for $\mathcal Z$. The
original work on autonomous flows \cite{Muell09} relied on an
imaginary part $\eta$ of the quasi-energy variable much larger
than a mean spacing; return to real quasi-energy was possible only
after summing up the series. The later extension to Floquet maps
\cite{Braun12b} could make do with 'infinitesimal' $\eta$. Our
present investigation reveals limits within which $\eta$ must lie
for the semiclassical periodic-orbit expansions to be applicable
to individual Floquet maps, at least for the correlator $R(e)$ in
the limit $N\to \infty$.

On the other hand, we have found here that one may stick to real
quasi-energies (in the sense $\eta\downarrow 0$) if one smoothes
by going to the first primitive $R^{(1)}(e)$; self-averaging then
takes place for $N\to\infty$ and even for large finite $N$ apart
from negligible noise. The semiclassical periodic-orbit expansions
can be done under the protection of that smoothing, and then
smoothing by integration also 'has the right' to give the
single-dynamics $R^{(1)}(e)$ equal to the CUE average.

Inasmuch as periodic orbits yield the CUE average of the full
generating function $\mathcal Z(e,\delta_+,\delta_-)$ it appears
that 'too much is proven' since we now know that a single spectrum
comes with gross violations of self-averaging for large
$\delta_\pm$.  One does not need to worry too much since large
$\delta_\pm$ do not harbor any physically relevant information.
But with the correlator (or its primitive) satisfactorily treated
one can live with subjecting the periodic-orbit expansion to a
suitable further average to justify the RMT result for $\mathcal
Z$.  We have checked that an average over $N\propto{1\over\hbar}$
leaving unchanged the classical limit does the job for our kicked
top. Averages over small intervals of classical control parameters
(small in the sense of shrinking to zero length for $\hbar\to 0$)
have been shown to work as well \cite{Zirnb99}.

We conclude with a speculative outlook. As already
stated above and illustrated in Fig.~\ref{fig:gc}, the second
primitive enhances large-$e$ structures. Exploiting that property we
have averaged the primitive $R^{(2)}(e)$ for single kicked-top spectra
over a range $[j,j+\Delta j]$ with $1\ll \Delta j\ll j$. That average
did away with noise at large $e \propto N=2j+1$ but left oscillations
in $e$, around the CUE average $R_{\mathrm{CUE}}^{(2)}(e)$. On the
other hand, when doing the same with matrices randomly drawn from the
CUE we found no oscillations. Further investigation must reveal
whether the oscillations found for the top are due to Ehrenfest-time
effects \cite{Tian04,Brouw06,Waltn10}, short orbits, or some other
effect.

We thank Sven Gnutzmann for discussions and gratefully acknowledge
support by the Sonderforschungsbereich SFBTR12 ''Symmetries and
universality in mesoscopic systems'' of the Deutsche
Forschungsgemeinschaft.

\section{\protect\bigskip Appendix}

\subsection{RMT predictions for covariance and variance of the generating functions,
generalized correlators and correlator primitives}
\label{RMTcovvar}

The covariance of two generating functions $Z\left(
a_{i},b_{i},c_{i},d_{i}\right) \equiv Z\left( i\right) ,\quad
i=1,2,$ can be
defined as%
\[
\Cov\left\{ Z\left( 1\right) ,Z\left( 2\right) \right\} =\left\langle
Z\left( 1\right) \,Z\left( 2\right) \right\rangle -\left\langle Z\left(
1\right) \right\rangle \,\left\langle Z\left( 2\right) \right\rangle
\]%
where $\left\langle \dots \right\rangle $ stands for the CUE
average; the second summand in the right -hand side is a product
of the well-known CUE generating functions. The first summand is
formally a double integral over the central phases $\phi _{1,2}$
of the two DD/DD ratios, but in view of the CUE averaging only
integration over the relative central phase  is needed,
\begin{eqnarray*}
\left\langle Z\left( 1\right) \,Z\left( 2\right) \right\rangle
&=&\int_{0}^{2\pi }\frac{d\phi }{2\pi }\left\langle \frac{\det \left( 1-c_{1}%
\mathrm{e}^{\I\phi }U\right) \det \left\langle
1-d_{1}\mathrm{e}^{-\I\phi }U^{\dagger }\right\rangle }{\det
\left( 1-a_{1}\mathrm{e}^{\I\phi }U\right)
\det \left( 1-b_{1}\mathrm{e}^{-\I\phi }U^{\dagger }\right) }\right.  \\
&&\times \left. \frac{\det \left( 1-c_{2}U\right) \det \left\langle
1-d_{2}U^{\dagger }\right\rangle }{\det \left( 1-a_{2}U\right) \det \left(
1-b_{2}U^{\dagger }\right) }\right\rangle .
\end{eqnarray*}%
The integrand can be imported from \cite{Conre07}, Eq.(1.1);
integration over $\phi $ is straightforward. The result can be
reformulated as the covariance of two generalized correlators
(\ref{eq:4}).  We present the final expression only for the
special case when in both correlators $\delta _{+}=-\delta
_{-}=\delta $; denoting $z_{k}=\e^{\I 2e_{k/}/N},k=1,2,$ we have
then,
\begin{equation}
\Cov\left\{ C\left( e_{1},\delta \right) ,C\left( e_{2},\delta
\right) \right\} =\frac{4}{N^{4}}A\left( z_{1},z_{2},\delta \right)
\,\,B\left( z_{1},z_{2}\right) ,
\end{equation}%
with%
\begin{eqnarray*}
A\left( z_{1},z_{2},\delta \right)  &=&\frac{z_{1}z_{2}}{\left(
z_{1}z_{2}-1\right) ^{2}}\prod_{k=1,2}\frac{\left(
z_{k}\e^{\I\delta /N}-1\right) \left( z_{k}\e^{-\I\delta
/N}-1\right) }{\left( z_{k}-1\right)
^{2}}; \\
B\left( z_{1},z_{2}\right)  &=&-\frac{\,z_{1}z_{2}+1}{z_{1}z_{2}-1}+\frac{%
\,z_{1}z_{2}-1}{z_{1}-z_{2}}\left[ \frac{z_{2}^{N}\left( z_{1}+1\right) }{%
z_{1}-1}-\frac{z_{1}^{N}\left( z_{2}+1\right) }{z_{2}-1}\right]  \\
&&+z_{2}^{N}z_{1}^{N}\left( \frac{2}{z_{1}-1}+\frac{2}{z_{2}-1}+\frac{6}{%
z_{1}z_{2}-1}+3-2N\right)
\end{eqnarray*}%
Assuming $\delta $  real we have for the complex conjugated correlator, $%
\bar{C}\left( e,\delta \right) =$ $C\left( \bar{e},\delta \right) $.
Therefore the covariance of the correlator real parts can be written,%
\begin{eqnarray}
&&\Cov\left\{ \mathrm{Re\,}C\left( e_{1},\delta \right) ,\mathrm{Re\,}%
C\left( e_{2},\delta \right) \right\}   \label{recovar} \\
&=&\frac{1}{2}\mathrm{Re\,}\left[ \Cov\left\{ C\left( e_{1},\delta
\right) ,C\left( e_{2},\delta \right) \right\} +\Cov\left\{ C\left(
e_{1},\delta \right) ,C\left( \bar{e}_{2},\delta \right) \right\} \right]
\nonumber
\end{eqnarray}
The variance of the correlator real part is obtained from (\ref{recovar}) in
the limit $e_{1},e_{2}\rightarrow e,$
\begin{eqnarray}\label{varsmallimpart}
\Var_{\mathrm{CUE}}\mathrm{Re\,}C\left( e,\delta \right)  &\equiv
&\left\langle \left( \mathrm{Re\,}C\left( e,\delta \right) \right)
^{2}\right\rangle -\left\langle \mathrm{Re\,}C\left( e,\delta \right)
\right\rangle ^{2} \nonumber\\
&=&\frac{1}{2}\mathrm{Re\,}\left[ A\left( z,z,\delta \right) B\left( z,z\right)
+A\left( z,\bar{z},\delta \right) B\left( z,\bar{z}\right) \right] \, ;
\end{eqnarray}%
note that
\begin{eqnarray*}
B\left( z,z\right)  &=&-\frac{z^{2}+1}{z^{2}-1}-z^{N-1}\frac{\left(
z+1\right) \left[ 2z+N\left( z^{2}-1\right) \right] }{z-1} \\
&&+z^{2N}\left( \frac{4}{z-1}+\frac{6}{z^{2}-1}+3-2N\right) .
\end{eqnarray*}

In the limit $\mathrm{Im\,}e_{1,2}\downarrow 0$ the covariance of
the physical correlators $R\left( e_{1,2}\right)
=\mathrm{Re\,}C\left( e_{1,2}+\I 0,0\right) $
becomes a sum of two expressions differing by the replacement $%
e_{2}\rightarrow -e_{2},$%
\begin{eqnarray}
\Cov\left\{ R(e_{1}),R(e_{2})\right\}  &=&\left[ \delta \left(
e_{1}-e_{2}\right) \frac{\pi }{N}\left( 1-\frac{\sin ^{2}e_{1}}{N^{2}\sin
^{2}\frac{e_{1}}{N}}\right) +\frac{\cos 2\left( e_{1}-e_{2}\right) }{%
N^{3}\sin ^{2}\frac{e_{1}-e_{2}}{N}}\right.   \label{recovar_delta0} \\
&&\left. +\frac{\sin 2\left( e_{1}-e_{2}\right) \left( 1+\cos \frac{2e_{1}}{N%
}+\cos \frac{2e_{2}}{N}-3\cos \frac{2\left( e_{1}-e_{2}\right) }{N}\right) }{%
8N^{4}\sin ^{3}\frac{e_{1}-e_{2}}{N}\sin \frac{e_{1}}{N}\sin \frac{e_{2}}{N}}%
\right]   \nonumber \\
&&+\left[ \quad e_{2}\rightarrow -e_{2}\quad \right] ;  \nonumber
\end{eqnarray}%
the variance becomes infinite as signalled by $\delta \left(
e_{1}-e_{2}\right) $.

The covariance of the real parts of two generalized correlators
with $e_{1,2}>0$ is proportional to (\ref{recovar_delta0}),
\begin{eqnarray}
&&\Cov\left\{ \mathrm{Re\,}C\left( e_{1},\delta \right) ,\mathrm{Re\,}%
C\left( e_{2},\delta \right) \right\}   \label{recovar_delta_ne_0} \\
&=&\left( 1-\frac{\sin ^{2}\frac{\delta }{2N}}{\sin ^{2}\frac{e_{1}}{N}}%
\right) \left( 1-\frac{\sin ^{2}\frac{\delta }{2N}}{\sin ^{2}\frac{e_{2}}{N}}%
\right) \Cov\left\{ R(e_{1}),R(e_{2})\right\} ;  \nonumber
\end{eqnarray}%
in addition, it contains singular terms $\propto \delta \left(
e_{1}\right),\, \delta \left(
e_{2}\right)  $ which we do not write out.

The variance of the real part of the correlator first primitive
$C^{\left( 1\right) }\left( e,\delta \right) =\int_{e}^{N\pi
/2}de^{\prime }C\left(
e^{\prime },\delta \,\right) $ is obtained from the covariance (\ref{recovar}%
) by double integration,
\begin{eqnarray*}
&&\Var\left\{ \mathrm{Re\,}C^{\left( 1\right) }\left( e,\delta \right)
\right\} \equiv \left\langle \mathrm{Re\,}C^{\left( 1\right) }\left( e,\delta
\,\right) \mathrm{Re\,}C^{\left( 1\right) }\left( e,\delta \right) \right\rangle
-\left\langle \mathrm{Re\,}C^{\left( 1\right) }\left( e,\delta \right)
\right\rangle ^{2} \\
&=&\int_{e}^{N\pi /2}\int_{e}^{N\pi /2}\Cov\left\{ \mathrm{Re\,}%
C\left( e\,_{1},\delta \right) ,\mathrm{Re\,}C\left( e_{2}\,,\delta \right)
\right\} de_{1}de_{2}.
\end{eqnarray*}%
In view of the relation $\int_{0}^{e}\mathrm{Re\,}C\left(
e^{\prime },\delta \right) de^{\prime }=-\int_{e}^{N\pi
/2}\mathrm{Re\,}C\left( e^{\prime },\delta \right) de^{\prime }$
the integration area can also be the square $[0\leq
e_{1,2}\leq e]$; singularities of the covariance at $e_{1,2}=0$ omitted in (%
\ref{recovar_delta_ne_0}) have then to be taken into account.

\subsection{Variance of generalized correlator in CUE, COE and CSE: numerical simulation}
\label{numercuecoe} 
\begin{figure}
[h]
\begin{center}
\includegraphics[
scale=0.35
]%
{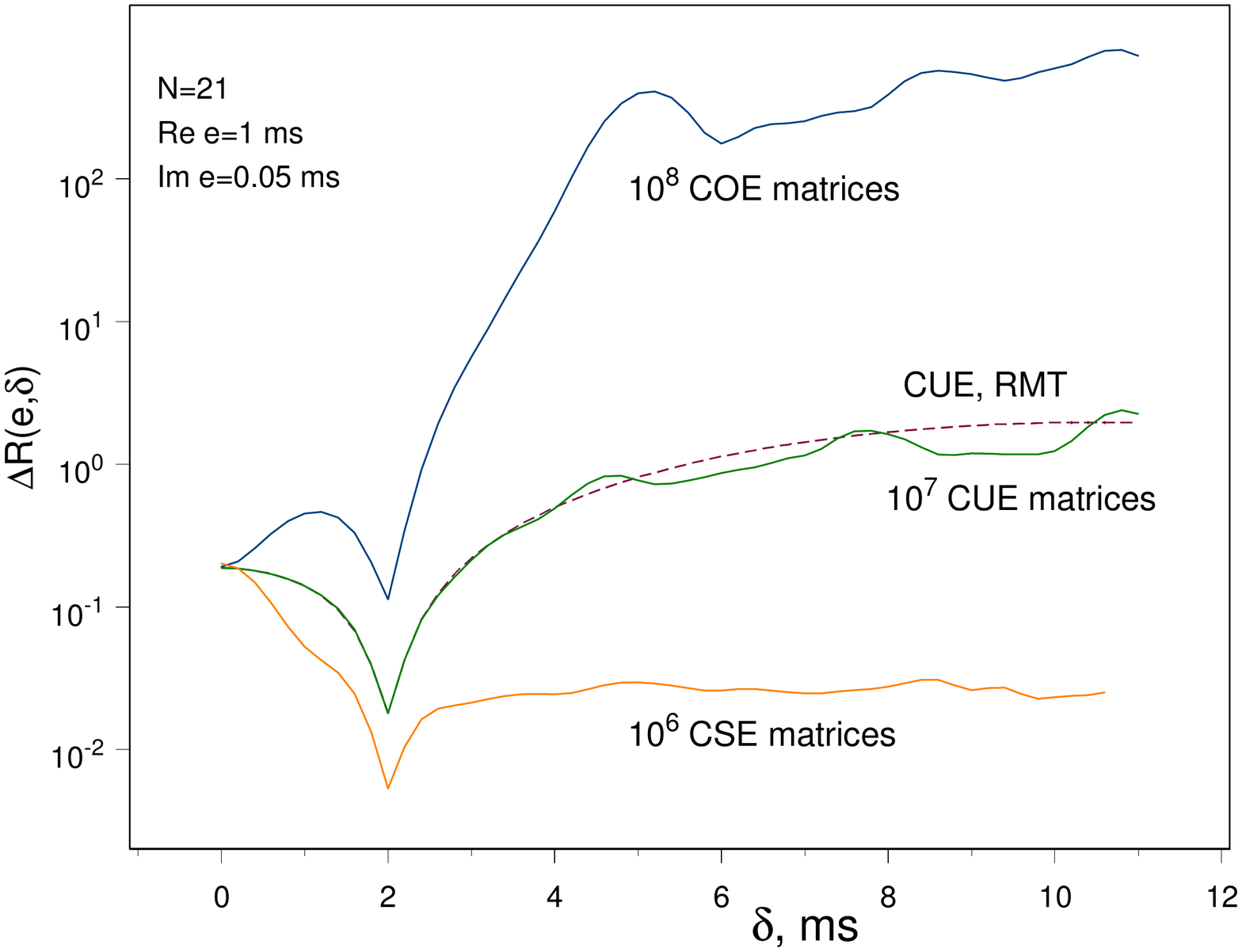}%
\caption{Deviation of generalized correlator regularized by
$\mathrm{Im}\,e=0.05$ ms. Numerical simulation with $10^{8}$
(COE), $10^{7}$ (CUE) and $10^{6}$ (CSE) matrices;  dashed line
depicts RMT prediction for
CUE}%
\label{COE}%
\end{center}
\end{figure}

As we pointed out in the main text, the
variance of the generalized correlator as function of $\delta$
must be drastically different for the three Dyson ensembles
because of different level repulsion: Compared with CUE, $\beta=2,
$ the correlator fluctuations at large $\delta$ must be more
pronounced in COE, $\beta=1,$ and less so in CSE, $\beta=4$. We
checked these predictions computing the standard deviation as
function of $\delta$ in the ensembles of $10^{7}$ CUE, $10^{6}$
CSE and $10^{8}$ COE matrices; the CSE matrices were generated by
the method suggested in \cite{Zyczk95}. The chosen number of
matrices in each ensemble depended on the rate of convergence
which was the slowest in COE and the fastest in CSE. The real part
of energy was fixed at $e_{0}=1$ ms and its imaginary part at
$\eta=0.05$ ms; the matrix size was $N=21$. The result is shown by
the full curves in Fig. \ref{COE}.

The deviation in the three ensembles was close at $\delta=0$, i.
e., for the physical correlators. On the other hand, at large
$\delta$ the mean deviations of the numerical COE and CSE
correlator differed, in opposite directions, from its CUE
counterpart by up to three orders of magnitude; this can be
interpreted as confirmation of our hypothesis on the role of the
level repulsion.

The dashed curve in Fig. \ref{COE} depicts the theoretical CUE\ deviation
following from the exact Eq. (\ref{varsmallimpart}); good agreement with
the numerics in a wide range is obvious.

Note the sharp minimum of the deviation at $\delta=2e_{0}$. It is
predicted in the CUE case by
  (\ref{vargencorrsmallim}), exact in the limit $\mathrm{Im}\,e\to +0$. The zero deviation
indicates that the generalized correlator $C\left( e_{0}+\I
0,2e_{0}\right)  $ of any individual spectrum is the same. This is
indeed the case; moreover, that spectrum-independent function
coincides with the CUE correlator.
\newpage
\section*{References}

\bibliographystyle{unsrt}
\bibliography{pbraun6}

\end{document}